%% file: main.tex
\documentclass[sigconf,anonymous=False]{acmart}
\usepackage{multirow}
\usepackage{bm}
\usepackage{pgfplots}
\usepackage{caption}
\usepackage{subcaption}

\usepackage[ruled]{algorithm2e}
\usepackage{algorithmic}
\AtBeginDocument{%
  \providecommand\BibTeX{{%
    \normalfont B\kern-0.5em{\scshape i\kern-0.25em b}\kern-0.8em\TeX}}}

\copyrightyear{2022}
\acmYear{2022}
\setcopyright{acmlicensed}\acmConference[SIGIR '22]{Proceedings of the 45th International ACM SIGIR Conference on Research and Development in Information Retrieval}{July 11--15, 2022}{Madrid, Spain}
\acmBooktitle{Proceedings of the 45th International ACM SIGIR Conference on Research and Development in Information Retrieval (SIGIR '22), July 11--15, 2022, Madrid, Spain}
\acmPrice{15.00}
\acmDOI{10.1145/3477495.3531813}
\acmISBN{978-1-4503-8732-3/22/07}

\begin{document}

%\title{Learning disentangled item representations for recommendation via Momentum Contrast with Pointwise and Pairwise Labels}
\title{MP2: A Momentum Contrast Approach for Recommendation with Pointwise and Pairwise Learning}
% \author{Menghan Wang}
% \email{menghawang@ebay.com}
% \affiliation{%

\author{Menghan Wang} 
\email{wangmengh@zju.edu.cn}
\affiliation{%
  \institution{eBay Inc.} 
  \city{Shanghai}
  \country{China} 
}
\author{Yuchen Guo} 
\email{yuchguo@ebay.com}
\affiliation{%
  \institution{eBay Inc.} 
  \city{Shanghai}
  \country{China} 
}
\author{Zhenqi Zhao} 
\authornote{Work done while at eBay.}
\email{kyriezhao@tencent.com}
\affiliation{%
  \institution{Tencent Inc.} 
  \city{Shanghai}
  \country{China} 
}
\author{Guangzheng Hu} 
\email{Guangzhengh@student.unimelb.edu.au}
\affiliation{%
  \institution{The University of Melbourne} 
  \city{Melbourne}
  \country{Australia} 
}
\author{Yuming Shen} 
\authornotemark[1]
\email{yuming.shen@eng.ox.ac.uk}
\affiliation{%
  \institution{University of Oxford} 
  \city{London}
  \country{United Kingdom} 
}
\author{Mingming Gong} 
\email{mingming.gong@unimelb.edu.au}
\affiliation{%
  \institution{The University of Melbourne} 
  \city{Melbourne}
  \country{Australia} 
}
\author{Philip Torr} 
\email{philip.torr@eng.ox.ac.uk}
\affiliation{%
  \institution{University of Oxford} 
  \city{London}
  \country{United Kingdom} 
}
 \renewcommand{\shortauthors}{Wang, et al.}


%\renewcommand{\shortauthors}{Anonymous authors} 
\newcommand{\method}{MP2}
\fancyhead{} 
\begin{abstract}
Binary pointwise labels (aka implicit feedback) are heavily leveraged by deep learning based recommendation algorithms nowadays. In this paper we discuss the limited expressiveness of these labels may fail to accommodate varying degrees of user preference, and thus lead to conflicts during model training, which we call annotation bias. To solve this issue, we find the soft-labeling property of pairwise labels could be utilized to alleviate the bias of pointwise labels. To this end, we propose a momentum contrast framework (\method{}) that combines pointwise and pairwise learning for recommendation. \method{} has a three-tower network structure: one user network and two item networks. The two item networks are used for computing pointwise and pairwise loss respectively. To alleviate the influence of the annotation bias, we perform a momentum update to ensure a consistent item representation. Extensive experiments on real-world datasets demonstrate the superiority of our method against state-of-the-art recommendation algorithms. 
%Deep learning based recommendation algorithms have become prevalent nowadays, most of which are pointwise methods with binary labels. However, user preferences are of varying degrees, and we argue that the limited expressiveness of binary pointwise labels cannot fully reveal the true preference of users towards items, leading to contradictory optimizing signals and unstable representations during model training, which we call annotation bias. To solve this issue, we find pairwise labels are soft labeling and free of the annotation bias, and could be leveraged to correct the bias of pointwise labels. In this paper, we propose a momentum contrast framework (\method{}) that combines pointwise and pairwise learning for recommendation. \method{} has a three-tower network structure: one user network and two item networks. The two item networks are used for computing pointwise and pairwise loss respectively. To alleviate the influence of the annotation bias, we perform a momentum update to ensure a consistent item representation. Besides, MP2 automatically tune the weights of pointwise labels with discrepancy between representations generated by two networks. Extensive experiments on real-world datasets demonstrate the superiority of our method against state-of-the-art recommendation algorithms. 
\end{abstract}


% \ccsdesc[500]{Information systems~Collaborative filtering}



\maketitle

\input{intro}
\input{model}
\input{experiments}
\input{conclusion}

\section{Acknowledgements}
Yuming Shen is partially supported by the UKRI grant: Turing AI Fellowship EP/W002981/1 and EPSRC/MURI grant: EP/N019474/1. Yuming also acknowledge the philanthropic support of the donors to the University of Oxford's COVID-19 Research Response Fund: BRD00230. Yuming would like to thank the Royal Academy of Engineering and FiveAI. The authors thank the anonymous reviewers for their helpful comments.

%\section{Acknowledgments}

%\begin{acks} 
%\end{acks}

\bibliographystyle{ACM-Reference-Format}
\bibliography{p2lr}
 
\end{document}

%% file: intro.tex
\section{Introduction}
\label{sec:intro} 
Personalized recommendation is becoming a key component in web applications. It is often cast as a learning-to-rank (LTR) problem~\citep{cheng2016wide,guo2017deepfm,naumov2019deep}, where an ordered list of items are selected to meet users' interest. 
% Recent years have seen a rapid development of recommendation algorithms, from linear regression~\citep{weisberg2005applied} to decision trees~\citep{chen2016xgboost}, and to deep models~\citep{cheng2016wide,naumov2019deep}. 
Due to the superior ability to learn from big datasets, deep learning-based recommendation algorithms are becoming the mainstream solution for practitioners.
 
 \begin{figure}[t]
\centering 
\includegraphics[width=.35\textwidth]{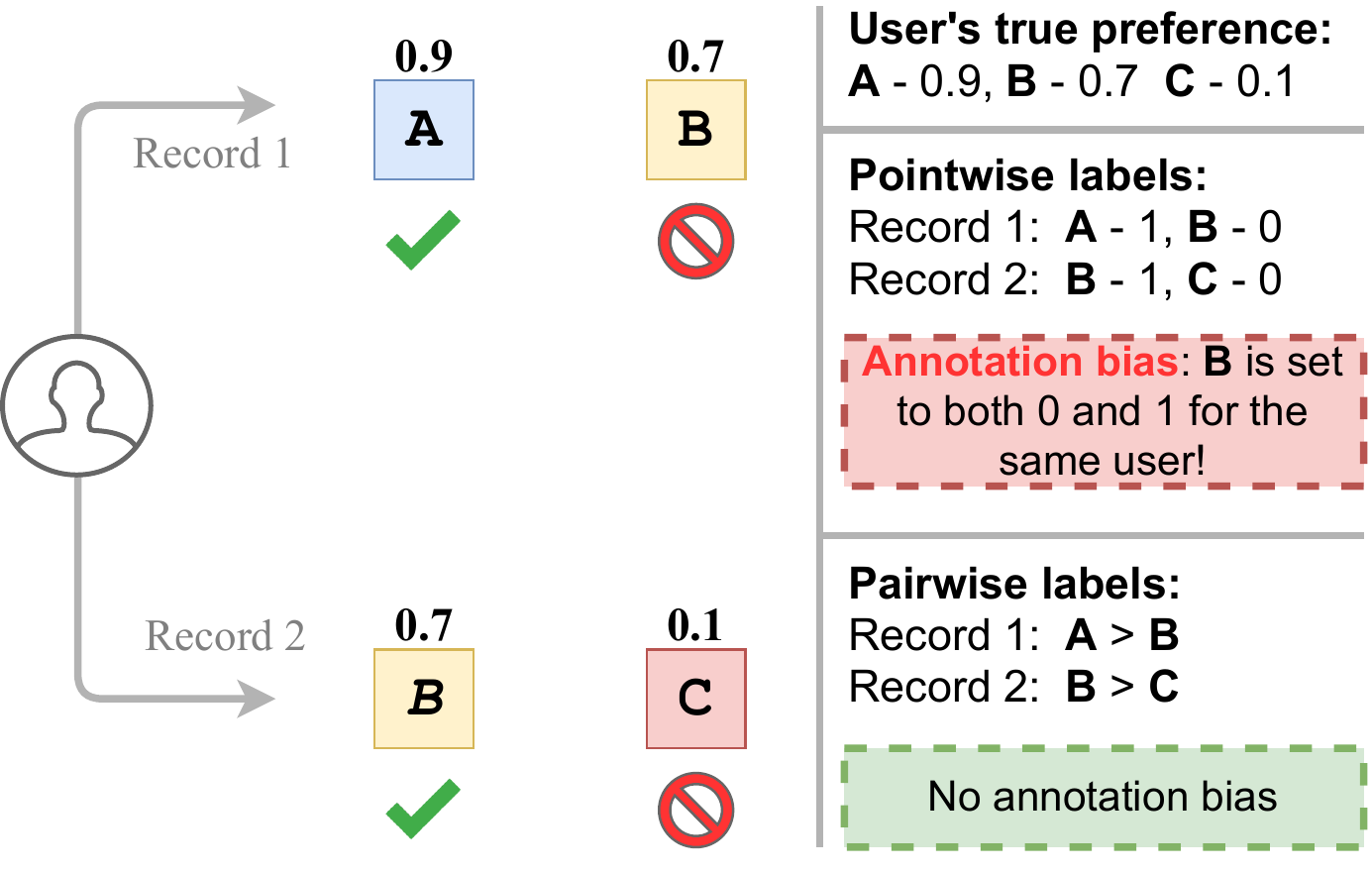}
\caption{An example of one user's two interaction records and the corresponding pointwise and pairwise labels. The pointwise labels of item $B$ have an annotation bias problem.}
\label{example}
\end{figure}

Since the implicit feedback (e.g., click or not click, purchase or not purchase) is abundant, binary labels are widely chosen in practice. However, user preferences are not stable and often influenced by context. For example, in Fig. \ref{example} we can see that item $B$ is annotated as $0$ and $1$ labels in different contexts but neither of them can accurately represent the user's exact preference $0.7$. We call it \textbf{annotation bias}, which is widely ignored in binary pointwise labels. (In reality, zero labels are also collected from users' impression logs, so the annotation bias is different from the well-studied exposure bias~\citep{wang2018collaborative,wang2018modeling}).
One drawback of having annotation bias for the deep model is a high fluctuation of learned representations. For example, the two conflict labels of item $B$ will give opposite optimizing signals to its item representation. 

On the other hand, pairwise labels are free of the annotation bias. As shown in Fig. \ref{example}, pairwise labels depict preference orders between items; it is a form of soft labeling and avoids annotating binary scores to items. Although pairwise learning is seldom studied in deep recommendations, traditional LTR studies have shown its effectiveness in learning users' comparative preferences. We argue that pointwise and pairwise learning are complementary to each other, and we could combine them to address the annotation bias. 
Particularly, in this paper, we propose a momentum contrast framework (\method{}) with pointwise and pairwise learning for recommendation. 
\method{} consists of a three-tower network structure: one user network and two item networks. The two item networks are used for computing pointwise and pairwise loss, respectively. \method{} also take two strategies to alleviate the annotation bias: momentum update and weighting label with discrepancy. The momentum update is applied to ensure a consistent item representation for pointwise learning, and weighting label with discrepancy aims to tune the weights of pointwise labels automatically. Extensive experiments we show that MP2 achieves
state-of-the-art performance compared with other competi-
tive algorithms.

\section{Related Work}
 \textbf{Deep Neural Networks in Recommendation.} In recent years, deep neural networks (DNNs) have become successful in recommendations, some representative examples are PNN \citep{qu2016product}, Wide\&Deep \citep{cheng2016wide}, DeepFM \citep{guo2017deepfm}, and DLRM \citep{naumov2019deep}. However, these methods fall into pointwise learning, leaving pairwise and listwise learning (beyond the scope of this work) almost blank in deep recommendations. There also exists researches \citep{chen2015fusing,wang2016ppp,lei2017alternating,cinar2020adaptive} combining pointwise and pairwise learning, most of which focuses on designing a mixed loss function. Instead in this paper, we focus on the backbone design.

%For example, \citet{lei2017alternating} formulates an adaptive strategy as minimizing a particular objective function that generalizes the traditional pointwise and pairwise loss function. \citet{cinar2020adaptive} proposes a joint learning model (APPL) that alternately draw pairwise and pointwise data samples, optimizing the respective losses. Different from the previous studies, we take the model structure and loss functions (i.e., pointwise and pairwise losses) in a unified perspective.

\textbf{Representation Learning with Momentum}
Momentum-based methods~\cite{memorybank,moco,mocov2} are intensively employed in the recent study in deep representation learning, of which the majority approaches require a set of \textit{slowly-progressing} parameter counterparts, updated with a momentum as reference during training. This idea has been proven to be fully functioning and effective in the context of self-supervised feature learning and pre-training in computer vision.  
However, we clarify that our proposed model differs from the aforementioned methods in motivation. Existing methods basically resort to the momentum-based approaches as a memory-saving solution to construct contrastive samples for comparison \cite{memorybank}, while we consider a slower evolving intensity with momentum on the item representations than the user ones throughout the training process for better performance.

%% file: model.tex
\section{MP2 Framework}
In this section we describe the proposed framework in details.
% We first depict the overall architecture of \method{}, followed by the three-tower design and its implementation. Then we discuss about the main properties of \method{}. At last, we introduce the model inference of \method{}.

% \subsection{Overall Architecture}
% The structure of \method{} is illustrated in Figure \ref{mainmodel}. From a high-level perspective, \method{} follows a three-tower design horizontally and applies an Embedding + MLP structure vertically. We inherit the setting of Embedding layer and MLP layer discussed in Sec \ref{embed+mlp}, which have shown superiority in dealing with continuous and categorical features for recommendation. we will skip it here. Below we mainly focus on the three-tower design and its implementation.

\begin{figure}[t]
\centering 
\includegraphics[height=6.5cm]{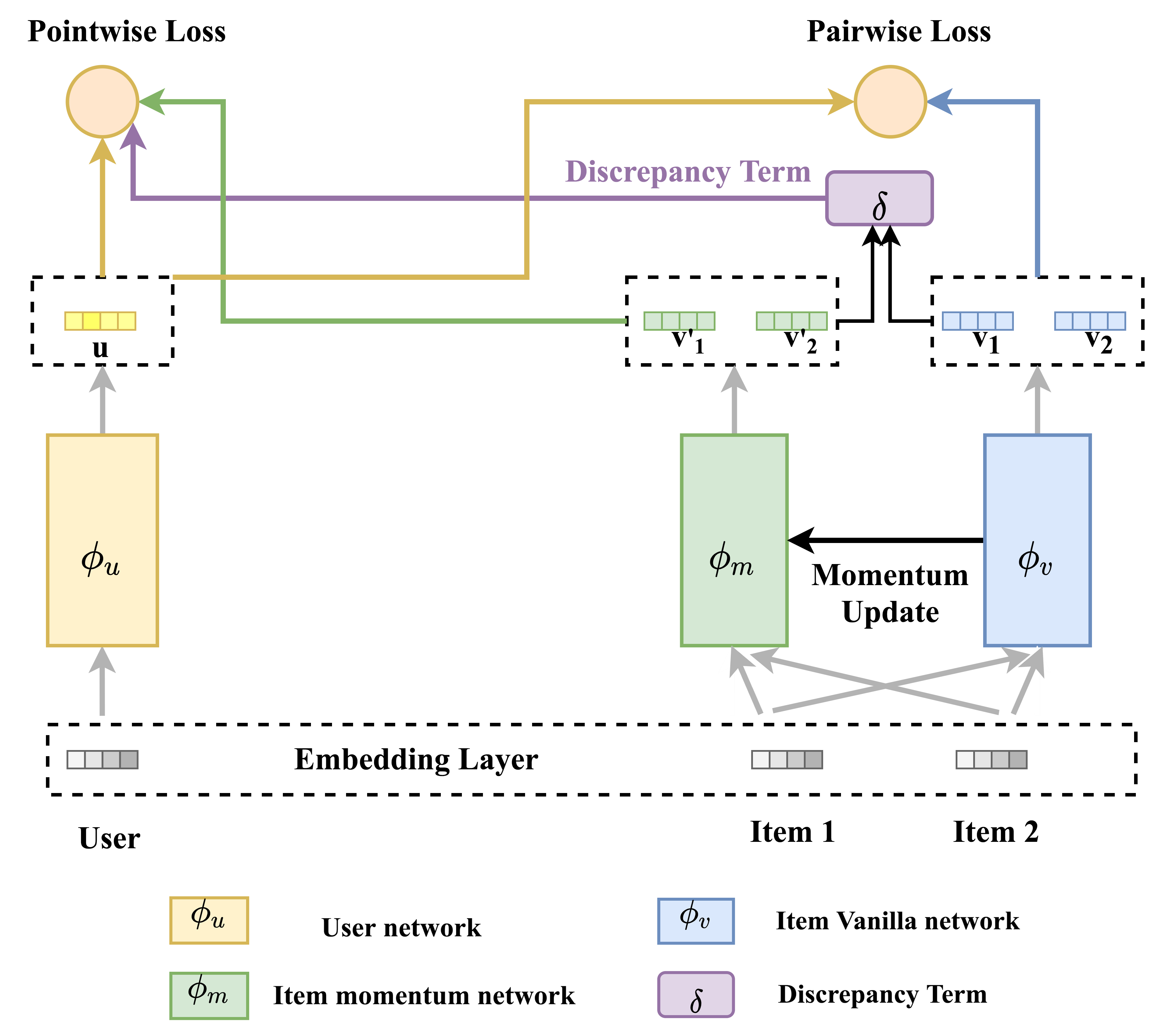}
\caption{Graphical architecture of \method{}. }
\label{mainmodel}
\end{figure}

\subsection{Three-Tower Design} 
From Fig. \ref{mainmodel} we can see a three-tower architecture; it consists of a user network $\phi_{u}(\cdot, \theta_{u})$, an item vanilla network $\phi_{v}(\cdot, \theta_{v})$, and an item momentum network $\phi_{m}(\cdot, \theta_{m})$, of which $\phi_{v}(\cdot, \theta_{v})$ and $\phi_{m}(\cdot, \theta_{m})$ have the same structure. Below the three towers are a feature embedding layer that deals with numerical and categorical features. The three towers use input feature embeddings to generate compact representations. Let $u$ be the user representation generated by $\phi_{u}(\cdot, \theta_{u})$, $v$ be the item representation generated by $\phi_{v}(\cdot, \theta_{v})$, and $v^{m}$ be the item representation generated by $\phi_{m}(\cdot, \theta_{m})$. We then use these representations for pointwise and pairwise learning. For a data sample $(U_i,V_j,V_t,y_{ij},y_{it},j>_{u}t)$, we use $\hat{y}_{ij}=u_iv_j^m$ and $\hat{y}_{it}=u_iv_t^m$ to predict pointwise labels, and use $\hat{y}_{pair}=u_iv_j-u_iv_t$ to predict pairwise labels. 

% Apart from the combination of pointwise and pairwise losses, the main difference between \method{} and other recommendation models is that we generate two representations for an item. Concretely, $\phi_{m}(\cdot, \theta_{m})$ generates item representation for pointwise learning, with $\phi_{v}(\cdot, \theta_{v})$ for pairwise learning.
Recalling that we aim to learn a consistent representation for items and this is the reason we design two item networks for representation learning. Intuitively, if the real-value of one item's representation changes a lot during the optimization, it indicates that the item representation fluctuates a lot and there may be an annotation bias in the corresponding labels. 
% In order to measure the fluctuation, a straightforward way is to store item representations in each optimization step and then compare them with those in the next step. However, it is unfeasible because of the requirement of huge storage memory when training on large-scale datasets. 

We leverage the two item networks to model the fluctuation and then address the annotation bias, which consists of two phases:
1) \textbf{\emph{Momentum update}}. Since pointwise labels suffer from an annotation bias, $v^m$ may highly fluctuate in traditional gradient-descent style optimizers. The item representation $v^m$ (pointwise learning) is optimized via momentum update rather than normal gradient back-propagation, which ensures a consistent, slowly-evolving update.
2) \textbf{\emph{Weighting label with discrepancy}}. The fluctuation is measured via the discrepancy of two representations of a same item, which is further served as the confidence term of the pointwise label. A higher fluctuation indicates a lower weight of the corresponding pointwise label, which automatically lower the importance of untrustworthy pointwise labels. 
% The implementation details of the two phases are shown below: 

\subsection{Momentum Update}

Different from the item vanilla network $\phi_{v}(\cdot, \theta_{v})$ that updates its weights $\theta_{v}$ via gradient back-propagation, $\phi_{m}(\cdot, \theta_{m})$ updates $\theta_{m}$ by averaging $\theta_{v}$:
\begin{equation}\label{eq:moco}
\theta_{m} = \alpha \theta_{m} + (1 - \alpha) \theta_{v},
\end{equation}
where $\alpha\in[0, 1)$ is a momentum coefficient hyper-parameter, whose value controls the smoothness of $\theta_{m}$. The momentum update in Eq. \ref{eq:moco} makes $\theta_{m}$ evolve more smoothly than $\theta_{v}$. As a result, though some items may have $0$ and $1$ pointwise labels for a same user due to the annotation bias, the fluctuation of item representations can be made small. In contrast, in a classical recommendation model ($y_{ij}=u_iv_j$) the $v_j$ will receive opposite optimizing signals, which will influence the consistency of item representations. 
% On the other hand, we encourage a fast evolving of user representation $u_i$ as user preference is dynamic and we want to capture the fluctuation.

The learning objective directly communicates $\phi_u\left(\cdot,\theta_u\right)$ through $\nabla_{u}L_{total}$ by back-propagation. For each step, the update of parameters $\theta_u$ then is instantly reflected in the user representations in the next batch. On the other hand, the item representations do not strictly follow this procedure with $\phi_v\left(\cdot, \theta_v\right)$. Instead, a momentum replicate $\phi_m\left(\cdot, \theta_m\right)$ processes all items. Heuristically, this results in temporally consistent item encoding to compile our motivation. 

\begin{algorithm}[t]
\small
\caption{The Training Procedure of MP2}
\label{alg}

\textbf{Input:}\hspace{0mm} A training dataset $\mathcal{D}=\{(U_i,V_j,V_t,y_{ij},y_{it},j>_{u}t)\}$.\\
\textbf{Output:}\hspace{0mm} Network parameters $\theta_v$, $\theta_v$ and $\theta_m$.\\
%	\BlankLine
%Randomly initialize $\mathbf{H}\in\{-1,1\}^{M\times N}$\\
\Repeat{convergence or reaching the maximum iteration}{
	Randomly select a mini-batch from $\mathcal{D}$\\
	$L_{total} \leftarrow$  Eq. \ref{eqn:final_loss}\\
	
	\textbf{Updating the vanilla and user network:}\\
	\hspace{4.3mm}$\theta_v\leftarrow\theta_v-\mathbf{\Gamma}\left(\nabla_{\theta_v}L_{total}\right)$\\
	\hspace{4.3mm}$\theta_u\leftarrow\theta_u-\mathbf{\Gamma}\left(\nabla_{\theta_u}L_{total}\right)$\\
	\textbf{Updating the momentum network:}\\
	\hspace{4.3mm}$\theta_{m} \leftarrow \alpha \theta_{m} + (1 - \alpha) \theta_{v}$ according to Eq.~\ref{eq:moco}\\
	
}

\end{algorithm}

\subsection{Weighting Label with Discrepancy}
After momentum update, we approximate the fluctuation with the discrepancy between two item representations $\phi_{v}(\cdot, \theta_{v})$ and $\phi_{m}(\cdot, \theta_{m})$. 
% That is, if item $j$'s representation changes a lot between $v_j$ and $v^m_j$, it means that the model has high uncertainty in learning the current label, and vice versa.
Formally, the discrepancy is defined as:
\begin{equation}\label{E_5}
    \bar{ \delta_j } =  \frac{1}{c} \sum_{d=1}^{c}( \delta_j^d),   \delta_j = |v_j - v^m_j|,
   \end{equation}

where $\delta_j$ is the element-level discrepancy and $||$ is an element-wise absolute value operation. $c$ is the vector length of $\delta_j$, and $\bar{ \delta_j } $ is a single value.
% where $||$ is an element-wise absolute value operation. 
% After getting the element-level discrepancy we can further obtain the item-level discrepancy by averaging all elements in $ \delta_j $:
% where $c$ is the vector length of $\delta_j$, and $\bar{ \delta_j } $ is a single value. 
We regard this whole discrepancy as the confidence term of the pointwise label. Intuitively, a large $\bar{ \delta_j }$ indicates a high uncertainty so we use its reciprocal as the confidence term of pointwise label. Note that we have two items (i.e., $j$ and $t$) in one data sample and they are related, we combine them for the labeling weight:
\begin{equation}\label{E_w}
 w_{jt}  =\frac{1}{\exp({  \bar{ \delta_j } +  \bar{ \delta_t }) }}.
\end{equation}

From Eq.~\ref{E_w}, the similarity of two representation replica from the corresponding stochastic transformations $\phi_v(\cdot,\theta_v)$ and $\phi_m(\cdot,\theta_m)$ reflects the locality of the trained item representation space, which 
ideologically coheres with temporal ensembling~\cite{pimodel} but with different approaches. \cite{pimodel} typically minimizes Eq.~\ref{E_5}, while we consider re-weighting the loss functions to relate this discrepancy with the user representations.

% (see Sec.~\ref{sec_441}). The locality of representations is of importance in describing the similarity of items. We show empirical results of this in Sec.~\ref{sec_56}.

\subsubsection{Loss Calculation}\label{sec_441}

As shown in Fig. \ref{mainmodel}, \method{} consists of two kinds of losses: pointwise loss and pairwise loss. For pointwise loss we use discrepancy term as the confidence to the data sample. 
\begin{equation}
L_{pointwise}  = - \displaystyle\sum_{j=1}^n   w_{jt} y_j \log(p_j) + w_{jt} (1-y_j) \log(1-p_j),
\label{eq:point_loss_new}
\end{equation}
where $p_j = \frac{1}{1 + \exp(-\hat{y}_j)}$. Note that $w_{jt}$ is the weight for both item $j$ and item $t$ in a data sample.
As for pairwise loss, we use the multiplication of user representation and item representation from momentum network to compute the pairwise.

\begin{equation}
L_{pairwise} = \displaystyle\sum_{u=1}^m\displaystyle\sum_{j=1}^n \displaystyle\sum_{t=1}^n  \mathbb{I}( j>_{u}t) \log(1 + \exp(u_i v_j^{\prime} > u_i v_t^{\prime}))),
\label{eq:pair_loss_new}
\end{equation}
where $\mathbb{I}(\cdot)$ is the indicator function.

Finally the total loss function becomes a linear combination of Eq. \ref{eq:point_loss_new} and Eq. \ref{eq:pair_loss_new}. We also add $L2$-regularization terms to avoid overfitting:
 \begin{equation}
\label{eqn:final_loss}
L_{total}= L_{pointwise}  +  \beta * L_{pairwise}  + regularization.
\end{equation}

%% file: experiments.tex
\section{Experiments}
\label{sec:exp} 
In this section, we conduct experiments to evaluate the effectiveness of \method{}. 
We compare \method{} with competitive baselines, including pointwise and pairwise algorithms. Following that, we conduct additional experiments for investigating effectiveness of each component in \method{}.
% Finally, we reveal the semantic discriminability of the learned item representation with a case study.

% \begin{table*}[h]
%     \small
% 	\centering
% 	\caption{Datasets Statistics.}
% 	\label{tb:statistics}
% 	\begin{tabular}{c|ccc|ccccc}
% 		\toprule
% 		Dataset      & \#User & \#Item  & \#Rating  & \vtop{\hbox{\strut Pointwise}\hbox{\strut Sparsity}} & \vtop{\hbox{\strut Pairwise}\hbox{\strut Sparsity}} & \vtop{\hbox{\strut Average positive}\hbox{\strut sample/user}}   & \vtop{\hbox{\strut Average negative}\hbox{\strut sample/user}}  & Positive Ratio     \\ \midrule
		
% 		Movielens-100k      & 943 & 1,682  & $1 \times 10^5$ &  6.30\%  & 12.07\% &  17.87 & 66.96 & 21.07\%    \\ \midrule
% 		Movielens-1M      & 6,040 & 3,706  & $1 \times 10^6$ &  4.46\%  &  5.85\% &  29.98 & 102.49 & 22.63\%    \\ \midrule
% 		Beauty      & 22,363  & 12,101  & $2 \times 10^5$ &  0.07\%  &  0.38\%   &  4.09 & 3.01 & 57.68\%    \\ \midrule
% 		Office Products  & 4,905  & 2,420  & 53,258   & 0.45\%  & 2.71\%  & 4.94  & 3.74 & 56.92\% \\   \bottomrule
% 	\end{tabular}
% \end{table*}

 \begin{table*}[th]
 %\small
 \footnotesize
        \caption{Comparisons of different models on four datasets.}
        \label{tab:offexp}
        \begin{tabular}{l|c|c|c|c|c|c|c|c}
            \toprule
            \multirow{2}{*}{Models} &  \multicolumn{4}{c|}{Movielens-100k}              &                \multicolumn{4}{c}{Movielens-1m}          \\
            \cline{2-9}
                           & HitRate@5 & HitRate@20 & NDCG@5 & NDCG@20 & HitRate@5 & HitRate@20 & NDCG@5 & NDCG@20 \\ 
            \midrule
           NeuMF           &  0.4709 & 0.8163 & 0.3311 & 0.4570                 & 0.2182 & 0.5034 & 0.1961 & 0.3002 \\
           BPR          & 0.4550 & 0.8182 & 0.3172 & 0.4526                & 0.4810 & 0.7194 & 0.4369 & 0.5241 \\
           Ranknet-NN        & 0.5949 & 0.8822 & 0.4415 &  0.5493                 & 0.4074 & 0.6939 & 0.3498  & 0.4627 \\
           APPL    & 0.5877 &  0.8851 & 0.4590 & 0.5510                 & 0.3881 & 0.6347 & 0.3447 & 0.4330 \\
          \midrule
           T3     & 0.5927 & 0.8824 & 0.4531 & 0.5491            &  0.4832 &  0.7109 & 0.4391 & 0.5217 \\
            
            MP2         & \textbf{0.5983} & \textbf{0.8947} & \textbf{0.4721} & \textbf{0.5604}                   & \textbf{0.4920} & \textbf{0.7270} & \textbf{0.4449} & \textbf{0.5314} \\ 
            
            \bottomrule
%--------
\toprule
            \multirow{2}{*}{Models} &  \multicolumn{4}{c|}{Beauty}              &                \multicolumn{4}{c}{Office Products }          \\
            \cline{2-9}
                           & HitRate@5 & HitRate@20 & NDCG@5 & NDCG@20 & HitRate@5 & HitRate@20 & NDCG@5 & NDCG@20 \\ 
            \midrule
           NeuMF           & 0.2347 & 0.9038 & 0.1136 & 0.2616                 & 0.2229 & 0.8150 & 0.1145 & 0.2706 \\
           BPR          & 0.2555 & 0.8629 & 0.1275 & 0.2605                & 0.2754 & 0.8613 & 0.1504 & 0.2858 \\
           Ranknet-NN        & 0.3157 & 0.8824 & 0.1643 & 0.2889                    & 0.2923 & 0.8449 & 0.1537  & 0.2829 \\
           APPL    & 0.2780 & 0.8664 & 0.1429 & 0.2716                 & 0.2311 &  \textbf{0.8947} & 0.1192 & 0.2756 \\
          \midrule 
           T3     & 0.2811 & \textbf{0.9078} & 0.1440 & 0.2821             & 0.2657 & 0.8789 & 0.1436 & 0.2841 \\
            
            MP2         & \textbf{0.3170} & 0.8850 & \textbf{0.1668} & \textbf{0.2917}                   & \textbf{0.3031} & 0.8894 & \textbf{0.1586} & \textbf{0.2915} \\ 
          
            \bottomrule

        \end{tabular}
    \end{table*}

\subsection{Datasets and Experimental Settings}
In this section we introduce the used datasets and experimental settings, including baselines, offline metrics, and reproducibility.
\subsubsection{Datasets}
We selected four datasets for evaluating recommendation performance, two from MovieLens and two from Amazon.
The datasets from MovieLens\footnote{https://grouplens.org/datasets/movielens/} are ml-100k and ml-1m, separately. 
% The rating data sets are collected from MovieLens web site with different user and item sizes. 
The other two datasets are Beauty and Office Products, which collect product reviews and metadata from Amazon\footnote{http://snap.stanford.edu/data/amazon/productGraph/categoryFiles/}. The original ratings of the four datasets are explicit integer ratings range from $1$ to $5$.
% , and we transform them into two pointwise and pairwise labels.
For pointwise labels, we use specific threshold $3$ to binarize the rating scores as labels. For pairwise label, we select item pairs under the same user randomly, and then decide the labels based on relative scores of item pairs. 
% The construction of pairwise label is asymmetric, namely the label of ($y_j$, $y_t$) is different from ($y_t$, $y_j$).
% The statistics of the four datasets are summarized in Table~\ref{tb:statistics}.

\subsubsection{Baselines}
We evaluate the performance of \method{} against the following baseline models. Baseline models are chosen from three fileds: pointwise methods, pairwise methods, and pointwise+pairwise methods. 1) \textbf{NeuMF} \citep{he2017neural}. This is a neural network based collaborative filtering method with binary cross-entropy loss. It consists of a two-tower structure. 2) \textbf{BPR} \citep{rendle2012bpr}. This is a pairwise ranking method optimizing the matrix factorization model with a pairwise ranking loss, which is a classical pairwise recommendation model. 3) \textbf{Ranknet-NN} \citep{burges2005learning}.
  This is a neural network model applying pairwise loss and a two-tower structure.  
4) \textbf{APPL}. \citep{cinar2020adaptive} This model is a joint learning model that combines two pointwise losses and one pairwise loss. Its original version is based on matrix factorization, and we implemented a deep learning version that replaces matrix factorization with a two-tower neural network. 5) \textbf{T3 (Three-Tower)}. This model is a truncated version of \method{}, where we remove the momentum update and discrepancy term from \method{}. So this model contains a three-tower structure with two pointwise labels and one pairwise label.
Hyperparameter tuning is conducted by grid search, and each method is tested with the best hyperparameters for a fair comparison. 

\subsection{Performance Evaluation}\label{exp:performance}
We show the experimental results in Table \ref{tab:offexp}, from which we can find that \method{} outperforms other baselines consistently on each dataset. This demonstrates the effectiveness of our proposed method. 
Further, we can get the following findings. 
\textbf{1)} Models with joint loss (i.e., \method{}, T3, and APPL) are generally better than models with pairwise loss (Ranknet-NN and BPR) or pointwise loss (NeuMF), showing that combining pointwise and pairwise learning is a promising approach for recommendation. 
% Note that the data of \emph{Beauty} and \emph{Office Products} are extremely sparse, which lead to a deterioration of absolute values of metrics.
\textbf{2)} Pairwise models are empirically better than pointwise models. This is mainly because pairwise models capture relative relations of items and datasets are free of annotation bias.
\textbf{3)} Ranknet-NN (deep pairwise model) outperforms BPR (non-deep pairwise model) with a large margin on four datasets. Their loss function are the same and the difference is that Ranknet-NN applies a neural network, which could learn high-order feature interactions. In contrast, BPR is based on matrix factorization and it can only leverage shallow feature interactions for recommendation.
\textbf{4)} \method{} is superior to three tower and APPL, which indicates the effectiveness of momentum update and the weighting strategy. We also find that T3 is better than APPL, verifying the superiority of the three-tower structure over the two-tower structure. 
% As two item networks with the same structure may be redundant, we further put constrains on the network and the pointwise losses (T3 becomes \method{}), and achieve additional gains on rank metrics.  
By the above analysis, we can conclude that \method{} is effective and competitive.

% Recall that the loss function in eq. \ref{eqn:final_loss} is indeed a multi-task loss and a natural extension is to assign different weights to pointwise learning and pairwise learning, which is formulated as follows:
% \begin{equation}
% \label{eqn:final_loss_weighted}
% L_{total}= L_{pointwise}  +  \beta * L_{pairwise}  + regularization.
% \end{equation} 
% Note that discrepancy term in computing $L_{pointwise}$ functions as an instance-level weight and $\beta$ here is a task-level weight.
% We conduct another group of experiments to examine the trade off between pointwise and pairwise learning by varying $\beta$. From the Fig. \ref{exp:pairwise_weight} we can see that our model performs best when $\beta$ is around $0.6$. Note that the performance drops a lot when $\beta$ is $0$; in that case pairwise learning is not functioning, which shows the necessity to combine pointwise and pairwise loss.

\subsection{Effectiveness of Momentum Update} 
\method{} applies a momentum update strategy in the item momentum network in order to learn a consistent representation $\theta_{m}$. According to Eq. \ref{eq:moco}, the momentum coefficient hyper-parameter $\alpha\in[0, 1)$ controls the smoothness of $\theta_{m}$. To evaluate the effectiveness of the momentum coefficient  $\alpha$, we perform a grid search by varying $\alpha \in [0, 0.1, 0.5, 0.9, 0.99, 0.999, 0.9999, 1]$ to find the optimal value. A larger $\alpha$ means a slower update of $\theta_{m}$. Note that $\alpha =1$ indicates the item momentum network is equal to the item vanilla network at all times and there is no difference between $\theta_{m}$ and $\theta_{v}$. In other words, MP2 is deteriorated to a two-tower structure. Figure. \ref{exp:alpha} shows the model performance of \method{} with different $\alpha$. We can find that the performance of \method{} is increasing monotonously when  $\alpha$ increases, and reaches the peak when $\alpha=0.999$, which shows a smoother momentum update will yield a better item representation and thus improve the recommendation performance. These results also validate our assumption of a consistent item representation.

\begin{table}[th]
%\footnotesize
\small
\caption{Performance (NDCG) of different labeling weights of \method{}.}
\label{exp:discrepancy}
\begin{tabular}{l|ccc}
\toprule
%& uniform & static  & adapative \\ \midrule
 Movielens-100K&\method{}$_{uniform}$ & \method{}$_{separate}$  & \method{}$_{joint}$ \\ \midrule
NDCG@5& 0.4683&0.4702  & 0.4740 \\ \midrule
NDCG@20 &0.5640 &0.5617  & 0.5687 \\ \bottomrule
\toprule
Beauty&\method{}$_{uniform}$ & \method{}$_{separate}$  & \method{}$_{joint}$ \\ \midrule
NDCG@5& 0.1549&0.1519  & 0.1664 \\ \midrule 
NDCG@20 &0.2871 &0.2833 & 0.2897 \\ \bottomrule
\end{tabular}
\end{table}

\input{momentum_analysis_pic}

\subsection{Effectiveness of Discrepancy}

To evaluate the effectiveness of the discrepancy term, we compare our proposed method (denoted as \method{}$_{joint}$ in this subsection) against its two variants to: \method{}$_{uniform}$ with uniform weights and \method{}$_{separate}$ with separate weights for two pointwise labels in a data sample. Specifically, in a data sample $(U_i,V_j,V_t)$ the item $j$ is assigned a weight of $ w_{j}  =\frac{1}{\exp({  \bar{ \delta_j }) }}$, and the item $t$ has a weight of $ w_{t}  =\frac{1}{\exp({  \bar{ \delta_t}) }}$. Note that in \method{} the weights of two pointwise labels are both $w_{jt}  =\frac{1}{\exp({  \bar{ \delta_j } +  \bar{ \delta_t }) }}$. The only difference between \method{}$_{separate}$ and \method{}$_{joint}$ is whether the two pointwise labels of a data sample are used separately or jointly to compute the label weights. 
Table. \ref{exp:discrepancy} shows NDCH@5 and NDCG@20 of the three variants on two datasets: \emph{Movielens-100K} and \emph{Beauty}. We can see that \method{}$_{joint}$ is the best among the three variants, showing the effectiveness of its weighting strategy. We also find that \method{}$_{separate}$ is worse than \method{}$_{uniform}$; uniform weights seem to be more competitive. One reason is that \method{}$_{separate}$ is the most complex model and is prone to overfitting. Meanwhile, \method{}$_{separate}$ is more like "pointwise" weighting while \method{}$_{joint}$ has a form of "pairwise" weighting. \method{}$_{joint}$ considers relations between two pointwise labels and thus is more robust to the annotation bias.

\input{relation-loss-ndcg}

%% file: momentum_analysis_pic.tex
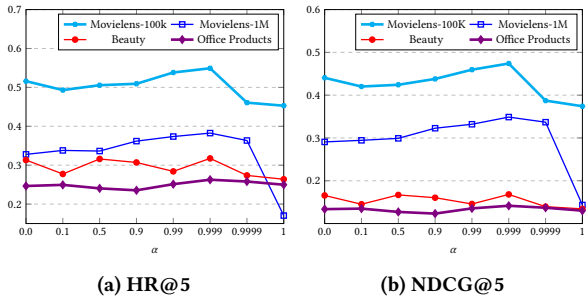
\begin{figure}
%--1st pic--
\raggedright
\subcaptionbox{HR@5 \label{fig:subfig18}}{%
%\begin{subfigure}[b]{0.2 \textwidth}
\raggedright
\begin{tikzpicture}[scale=0.5]
\begin{axis}[
    xlabel={$\alpha$},
    xmin=1, xmax=8,
    ymin=0.15, ymax=0.7,
    xtick={1,2,3,4,5,6,7,8},
    xticklabels={0.0,0.1,0.5,0.9,0.99,0.999,0.9999,1},
    %ytick={0.3,0.4,0.5,0.6,0.7,0.8 },
    %legend pos=north west,
    legend style={ legend columns=2},
    ymajorgrids=true,
    grid style=dashed,
]

    \addplot[
    color=cyan,
    mark=asterisk,
    line width=2pt,
    ]
    coordinates {
    (1,0.5159936658749009)(2,0.4930720506730007)(3,0.5056215360253362)(4,0.5092636579572446)(5,0.5382422802850358)(6,0.5491290577988917)(7,0.4607917656373714)(8,0.4528899445764054)
    };
   \addlegendentry{Movielens-100k}
     
    \addplot[
    color=blue,
    mark=square,
     line width=1pt,
    ]
    coordinates {
  (1,0.327805135)(2,0.337976316)(3,0.336408137)(4,0.361871263)(5,0.373657521)(6,0.382495017)(7,0.3634482)(8,0.170491851)

    };
     \addlegendentry{Movielens-1M}
     
 \addplot[
    color=red,
    mark=*,
    line width=1pt,
    ]
    coordinates {
    (1,0.313250767)(2,0.277443532)(3,0.315985134)(4,0.307033582)(5,0.284087961)(6,0.31747277)(7,0.273723928)(8,0.263884942)
%(1,0.142107)(2,0.135222)(3,0.123485)(4,0.135712)(5,0.107213)(6,0.116995)(7,0.13854) #verison1
    };
    \addlegendentry{Beauty}
     
 \addplot[
    color=violet,
    mark=diamond,
    line width=2pt,
    ]
    coordinates {
    (1,0.246519102)(2,0.249420605)(3,0.240412819)(4,0.235411009)(5,0.251217635)(6,0.262683324)(7,0.258039109)(8,0.249782727)
    };
   \addlegendentry{Office Products}

\end{axis}
\end{tikzpicture}
}
\subcaptionbox{NDCG@5 \label{fig:subfig2}}{%
%\end{subfigure}
%\hfil  \hfil  \hfil  
%\begin{subfigure}[b]{0.2  \textwidth}
\raggedleft
 \begin{tikzpicture}[scale=0.5]
\begin{axis}[
    xlabel={$\alpha$},
    xmin=1, xmax=8,
    ymin=0.1, ymax=0.6,
    xtick={1,2,3,4,5,6,7,8},
    xticklabels={0.0,0.1,0.5,0.9,0.99,0.999,0.9999,1},
    ytick={0.2,0.3,0.4,0.5,0.6,0.7,0.8 },
    %legend pos=outer north west,
    legend style={ legend columns=2},
    ymajorgrids=true,
    grid style=dashed,
]

    \addplot[
    color=cyan,
    mark=asterisk,
    line width=2pt,
    ]
    coordinates {
    (1,0.4405626911239629)(2,0.4201842319342088)(3,0.4242725017760578)(4,0.4379470710367412)(5,0.45951906836060125)(6,0.47399387638283397 )(7,0.3873084902818337)(8,0.37411131772280914)
    
    };
   \addlegendentry{Movielens-100K}
      
\addplot[
    color=blue,
    mark=square,
     line width=1pt,
    ]
    coordinates {
 (1,0.290748473)(2,0.294479451)(3,0.299099177)(4,0.32276111)(5,0.332032181)(6,0.3487884)(7,0.3368769)(8,0.143320765)

    };
     \addlegendentry{Movielens-1M}
     
 \addplot[
    color=red,
    mark=*,
    line width=1pt,
    ]
    coordinates {
     (1,0.165618626)(2,0.145123647)(3,0.166865602)(4,0.160383578)(5,0.145728174)(6,0.168056611)(7,0.139440674)(8,0.134243439) 
    };
    \addlegendentry{Beauty}
   
 \addplot[
    color=violet,
    mark=diamond,
    line width=2pt,
    ]
    coordinates {
     (1,0.133657521)(2,0.134884881)(3,0.127180272)(4,0.123253839)(5,0.135435851)(6,0.141565033)(7,0.136960116)(8,0.130082457)

    };
   \addlegendentry{Office Products} 
   
\end{axis}
\end{tikzpicture}    
%\end{subfigure}     
}
 \caption{ Performance of different momentum coefficient $\alpha$.}
\label{exp:alpha}
\end{figure}

%% file: conclusion.tex
\section{Conclusion}
\label{sec:conclusion}
In this paper, we study the annotation bias in recommendation, a widely existing but ignored problem caused by the limited expressiveness of binary pointwise labels.
We propose \method{}, a momentum contrast framework for recommendation that combines pointwise and pairwise learning to alleviate the annotation bias. 
%\method{} has a three-tower network structure: one user network and two item networks. The two item networks are used for computing pointwise and pairwise loss respectively. \method{} also take two strategies to alleviate the annotation bias: momentum update and weighting label with discrepancy. 
The offline experiments showed the superiority of \method{} over other competitive methods. In the future, we plan to combine listwise loss and pointwise loss in deep learning for recommendation.

%% file: main.bbl

\begin{thebibliography}{17}

%%% the Web address.

\ifx \showCODEN    \undefined \def \showCODEN     #1{\unskip}     \fi
\ifx \showDOI      \undefined \def \showDOI       #1{#1}\fi
\ifx \showISBNx    \undefined \def \showISBNx     #1{\unskip}     \fi
\ifx \showISBNxiii \undefined \def \showISBNxiii  #1{\unskip}     \fi
\ifx \showISSN     \undefined \def \showISSN      #1{\unskip}     \fi
\ifx \showLCCN     \undefined \def \showLCCN      #1{\unskip}     \fi
\ifx \shownote     \undefined \def \shownote      #1{#1}          \fi
\ifx \showarticletitle \undefined \def \showarticletitle #1{#1}   \fi
\ifx \showURL      \undefined \def \showURL       {\relax}        \fi
% invisible to TeX
\providecommand\bibfield[2]{#2}
\providecommand\bibinfo[2]{#2}
\providecommand\natexlab[1]{#1}
\providecommand\showeprint[2][]{arXiv:#2}

\bibitem[\protect\citeauthoryear{Burges, Shaked, Renshaw, Lazier, Deeds,
  Hamilton, and Hullender}{Burges et~al\mbox{.}}{2005}]%
        {burges2005learning}
\bibfield{author}{\bibinfo{person}{Chris Burges}, \bibinfo{person}{Tal Shaked},
  \bibinfo{person}{Erin Renshaw}, \bibinfo{person}{Ari Lazier},
  \bibinfo{person}{Matt Deeds}, \bibinfo{person}{Nicole Hamilton}, {and}
  \bibinfo{person}{Greg Hullender}.} \bibinfo{year}{2005}\natexlab{}.
\newblock \showarticletitle{Learning to rank using gradient descent}. In
  \bibinfo{booktitle}{\emph{ICML}}. \bibinfo{pages}{89--96}.
\newblock


\bibitem[\protect\citeauthoryear{Chen, Zhang, and Li}{Chen
  et~al\mbox{.}}{2015}]%
        {chen2015fusing}
\bibfield{author}{\bibinfo{person}{Lin Chen}, \bibinfo{person}{Peng Zhang},
  {and} \bibinfo{person}{Baoxin Li}.} \bibinfo{year}{2015}\natexlab{}.
\newblock \showarticletitle{Fusing pointwise and pairwise labels for supporting
  user-adaptive image retrieval}. In \bibinfo{booktitle}{\emph{ICMR}}.
  \bibinfo{pages}{67--74}.
\newblock


\bibitem[\protect\citeauthoryear{Chen, Fan, Girshick, and He}{Chen
  et~al\mbox{.}}{2020}]%
        {mocov2}
\bibfield{author}{\bibinfo{person}{Xinlei Chen}, \bibinfo{person}{Haoqi Fan},
  \bibinfo{person}{Ross Girshick}, {and} \bibinfo{person}{Kaiming He}.}
  \bibinfo{year}{2020}\natexlab{}.
\newblock \showarticletitle{Improved baselines with momentum contrastive
  learning}.
\newblock \bibinfo{journal}{\emph{arXiv preprint arXiv:2003.04297}}
  (\bibinfo{year}{2020}).
\newblock


\bibitem[\protect\citeauthoryear{Cheng, Koc, Harmsen, Shaked, Chandra, Aradhye,
  Anderson, Corrado, Chai, Ispir, et~al\mbox{.}}{Cheng et~al\mbox{.}}{2016}]%
        {cheng2016wide}
\bibfield{author}{\bibinfo{person}{Heng-Tze Cheng}, \bibinfo{person}{Levent
  Koc}, \bibinfo{person}{Jeremiah Harmsen}, \bibinfo{person}{Tal Shaked},
  \bibinfo{person}{Tushar Chandra}, \bibinfo{person}{Hrishi Aradhye},
  \bibinfo{person}{Glen Anderson}, \bibinfo{person}{Greg Corrado},
  \bibinfo{person}{Wei Chai}, \bibinfo{person}{Mustafa Ispir}, {et~al\mbox{.}}}
  \bibinfo{year}{2016}\natexlab{}.
\newblock \showarticletitle{Wide \& deep learning for recommender systems}. In
  \bibinfo{booktitle}{\emph{Proceedings of the 1st workshop on deep learning
  for recommender systems}}. \bibinfo{pages}{7--10}.
\newblock


\bibitem[\protect\citeauthoryear{Cinar and Renders}{Cinar and Renders}{2020}]%
        {cinar2020adaptive}
\bibfield{author}{\bibinfo{person}{Yagmur~Gizem Cinar} {and}
  \bibinfo{person}{Jean-Michel Renders}.} \bibinfo{year}{2020}\natexlab{}.
\newblock \showarticletitle{Adaptive Pointwise-Pairwise Learning-to-Rank for
  Content-based Personalized Recommendation}. In
  \bibinfo{booktitle}{\emph{Recsys}}. \bibinfo{pages}{414--419}.
\newblock


\bibitem[\protect\citeauthoryear{Guo, Tang, Ye, Li, and He}{Guo
  et~al\mbox{.}}{2017}]%
        {guo2017deepfm}
\bibfield{author}{\bibinfo{person}{Huifeng Guo}, \bibinfo{person}{Ruiming
  Tang}, \bibinfo{person}{Yunming Ye}, \bibinfo{person}{Zhenguo Li}, {and}
  \bibinfo{person}{Xiuqiang He}.} \bibinfo{year}{2017}\natexlab{}.
\newblock \showarticletitle{DeepFM: a factorization-machine based neural
  network for CTR prediction}. In \bibinfo{booktitle}{\emph{IJCAI}}.
  \bibinfo{pages}{1725--1731}.
\newblock


\bibitem[\protect\citeauthoryear{He, Fan, Wu, Xie, and Girshick}{He
  et~al\mbox{.}}{2020}]%
        {moco}
\bibfield{author}{\bibinfo{person}{Kaiming He}, \bibinfo{person}{Haoqi Fan},
  \bibinfo{person}{Yuxin Wu}, \bibinfo{person}{Saining Xie}, {and}
  \bibinfo{person}{Ross Girshick}.} \bibinfo{year}{2020}\natexlab{}.
\newblock \showarticletitle{Momentum contrast for unsupervised visual
  representation learning}. In \bibinfo{booktitle}{\emph{CVPR}}.
  \bibinfo{pages}{9729--9738}.
\newblock


\bibitem[\protect\citeauthoryear{He, Liao, Zhang, Nie, Hu, and Chua}{He
  et~al\mbox{.}}{2017}]%
        {he2017neural}
\bibfield{author}{\bibinfo{person}{Xiangnan He}, \bibinfo{person}{Lizi Liao},
  \bibinfo{person}{Hanwang Zhang}, \bibinfo{person}{Liqiang Nie},
  \bibinfo{person}{Xia Hu}, {and} \bibinfo{person}{Tat-Seng Chua}.}
  \bibinfo{year}{2017}\natexlab{}.
\newblock \showarticletitle{Neural collaborative filtering}. In
  \bibinfo{booktitle}{\emph{WWW}}. \bibinfo{pages}{173--182}.
\newblock


\bibitem[\protect\citeauthoryear{Laine and Aila}{Laine and Aila}{2016}]%
        {pimodel}
\bibfield{author}{\bibinfo{person}{Samuli Laine} {and} \bibinfo{person}{Timo
  Aila}.} \bibinfo{year}{2016}\natexlab{}.
\newblock \showarticletitle{Temporal ensembling for semi-supervised learning}.
\newblock \bibinfo{journal}{\emph{arXiv preprint arXiv:1610.02242}}
  (\bibinfo{year}{2016}).
\newblock


\bibitem[\protect\citeauthoryear{Lei, Li, Lu, and Zhao}{Lei
  et~al\mbox{.}}{2017}]%
        {lei2017alternating}
\bibfield{author}{\bibinfo{person}{Yu Lei}, \bibinfo{person}{Wenjie Li},
  \bibinfo{person}{Ziyu Lu}, {and} \bibinfo{person}{Miao Zhao}.}
  \bibinfo{year}{2017}\natexlab{}.
\newblock \showarticletitle{Alternating pointwise-pairwise learning for
  personalized item ranking}. In \bibinfo{booktitle}{\emph{ICDM}}.
  \bibinfo{pages}{2155--2158}.
\newblock


\bibitem[\protect\citeauthoryear{Naumov, Mudigere, Shi, Huang, Sundaraman,
  Wang, Gupta, Wu, Azzolini, et~al\mbox{.}}{Naumov et~al\mbox{.}}{2019}]%
        {naumov2019deep}
\bibfield{author}{\bibinfo{person}{Maxim Naumov}, \bibinfo{person}{Dheevatsa
  Mudigere}, \bibinfo{person}{Hao-Jun~Michael Shi}, \bibinfo{person}{Jianyu
  Huang}, \bibinfo{person}{Jongsoo Sundaraman}, \bibinfo{person}{Xiaodong
  Wang}, \bibinfo{person}{Udit Gupta}, \bibinfo{person}{Carole-Jean Wu},
  \bibinfo{person}{Alisson~G Azzolini}, {et~al\mbox{.}}}
  \bibinfo{year}{2019}\natexlab{}.
\newblock \showarticletitle{Deep learning recommendation model for
  personalization and recommendation systems}.
\newblock \bibinfo{journal}{\emph{arXiv preprint arXiv:1906.00091}}
  (\bibinfo{year}{2019}).
\newblock


\bibitem[\protect\citeauthoryear{Qu, Cai, Ren, Zhang, Yu, Wen, and Wang}{Qu
  et~al\mbox{.}}{2016}]%
        {qu2016product}
\bibfield{author}{\bibinfo{person}{Yanru Qu}, \bibinfo{person}{Han Cai},
  \bibinfo{person}{Kan Ren}, \bibinfo{person}{Weinan Zhang},
  \bibinfo{person}{Yong Yu}, \bibinfo{person}{Ying Wen}, {and}
  \bibinfo{person}{Jun Wang}.} \bibinfo{year}{2016}\natexlab{}.
\newblock \showarticletitle{Product-based neural networks for user response
  prediction}. In \bibinfo{booktitle}{\emph{ICDM}}. IEEE,
  \bibinfo{pages}{1149--1154}.
\newblock


\bibitem[\protect\citeauthoryear{Rendle, Freudenthaler, Gantner, and
  Schmidt-Thieme}{Rendle et~al\mbox{.}}{2012}]%
        {rendle2012bpr}
\bibfield{author}{\bibinfo{person}{Steffen Rendle}, \bibinfo{person}{Christoph
  Freudenthaler}, \bibinfo{person}{Zeno Gantner}, {and} \bibinfo{person}{Lars
  Schmidt-Thieme}.} \bibinfo{year}{2012}\natexlab{}.
\newblock \showarticletitle{BPR: Bayesian personalized ranking from implicit
  feedback}.
\newblock \bibinfo{journal}{\emph{arXiv preprint arXiv:1205.2618}}
  (\bibinfo{year}{2012}).
\newblock


\bibitem[\protect\citeauthoryear{Wang, Gong, Zheng, and Zhang}{Wang
  et~al\mbox{.}}{2018a}]%
        {wang2018modeling}
\bibfield{author}{\bibinfo{person}{Menghan Wang}, \bibinfo{person}{Mingming
  Gong}, \bibinfo{person}{Xiaolin Zheng}, {and} \bibinfo{person}{Kun Zhang}.}
  \bibinfo{year}{2018}\natexlab{a}.
\newblock \showarticletitle{Modeling dynamic missingness of implicit feedback
  for recommendation}.
\newblock \bibinfo{journal}{\emph{NeurIPS}}  \bibinfo{volume}{31}
  (\bibinfo{year}{2018}).
\newblock


\bibitem[\protect\citeauthoryear{Wang, Zheng, Yang, and Zhang}{Wang
  et~al\mbox{.}}{2018b}]%
        {wang2018collaborative}
\bibfield{author}{\bibinfo{person}{Menghan Wang}, \bibinfo{person}{Xiaolin
  Zheng}, \bibinfo{person}{Yang Yang}, {and} \bibinfo{person}{Kun Zhang}.}
  \bibinfo{year}{2018}\natexlab{b}.
\newblock \showarticletitle{Collaborative filtering with social exposure: A
  modular approach to social recommendation}. In
  \bibinfo{booktitle}{\emph{AAAI}}, Vol.~\bibinfo{volume}{32}.
\newblock


\bibitem[\protect\citeauthoryear{Wang, Wang, Tang, Liu, and Li}{Wang
  et~al\mbox{.}}{2016}]%
        {wang2016ppp}
\bibfield{author}{\bibinfo{person}{Yilin Wang}, \bibinfo{person}{Suhang Wang},
  \bibinfo{person}{Jiliang Tang}, \bibinfo{person}{Huan Liu}, {and}
  \bibinfo{person}{Baoxin Li}.} \bibinfo{year}{2016}\natexlab{}.
\newblock \showarticletitle{Ppp: Joint pointwise and pairwise image label
  prediction}. In \bibinfo{booktitle}{\emph{CVPR}}.
  \bibinfo{pages}{6005--6013}.
\newblock


\bibitem[\protect\citeauthoryear{Wu, Xiong, Yu, and Lin}{Wu
  et~al\mbox{.}}{2018}]%
        {memorybank}
\bibfield{author}{\bibinfo{person}{Zhirong Wu}, \bibinfo{person}{Yuanjun
  Xiong}, \bibinfo{person}{Stella~X Yu}, {and} \bibinfo{person}{Dahua Lin}.}
  \bibinfo{year}{2018}\natexlab{}.
\newblock \showarticletitle{Unsupervised feature learning via non-parametric
  instance discrimination}. In \bibinfo{booktitle}{\emph{CVPR}}.
  \bibinfo{pages}{3733--3742}.
\newblock


\end{thebibliography}
